\documentclass[11pt]{llncs}
\usepackage{graphicx}
\usepackage{amssymb}
\usepackage{listings}
\usepackage{pifont}
\usepackage{multirow}
\usepackage{url}
\usepackage{fullpage}
\newcommand{\keywords}[1]{\par\addvspace\baselineskip
\noindent\keywordname\enspace\ignorespaces#1}

\begin{document}

\title{Coarse-grained Dynamic Taint Analysis for Defeating Control and Non-control Data Attacks}

\urldef{\mailsa}\path|pankaj_kohli@research.iiit.ac.in|
\author{Pankaj Kohli}
\institute{Independent Researcher\\
New Delhi, India\\
\mailsa}

\maketitle


\begin{abstract}
Memory corruption attacks remain the primary threat for computer security. Information flow tracking or \textit{taint analysis} has been proven to be effective against most memory corruption attacks. However, there are two shortcomings with current taint analysis based techniques. First, these techniques cause application slowdown by about 76\% thereby limiting their practicality. Second, these techniques cannot handle non-control data attacks i.e., attacks that do not overwrite control data such as return address, but instead overwrite critical application configuration data or user identity data. In this work, to address these problems, we describe a coarse-grained taint analysis technique that uses information flow tracking at the level of application data objects. We propagate a one-bit taint over each application object that is modified by untrusted data thereby reducing the taint management overhead considerably. We performed extensive experimental evaluation of our approach and show that it can detect all critical attacks such as buffer overflows, and format string attacks, including non-control data attacks. Unlike the currently known approaches that can detect such a wide range of attacks, our approach does not require the source code or any hardware extensions. Run-time performance overhead evaluation shows that, on an average, our approach causes application slowdown by only 37\% which is an order of magnitude improvement over existing approaches. Finally, since our approach performs run-time binary instrumentation, it is easier to integrate it with existing applications and systems.

\keywords{Information Flow Tracking, Taint Analysis, Buffer Overflows, Format String Attacks, Non-control Data Attacks}

\end{abstract}

\section{Introduction}
\label{sec:intro}
Memory corruption attacks remain the primary threat for system breakins and security incidents. Vulnerabilities such as buffer overflows and format string attacks are used heavily to exploit applications and to allow attackers to gain illegitimate access. According to CERT, 14 out of 20 vulnerabilities with the highest severity metric are memory corruption vulnerabilities \cite{cert}. For remote applications, an important security concern is the nature of data that is being sent to the application from untrusted sources. Such untrusted data can cause damage, such as system break-ins, if it is not checked properly by the application. Hence, tracking the flow of such information to detect memory corruption attacks is a mandatory requirement to safeguard critical network applications. Signature based scanning is often too slow to respond to attacks such as the Microsoft Windows server service RPC handling remote code execution vulnerability (MS08-067) \cite{serversvc}, that was exploited by the Conficker worm \cite{conficker}. Such attacks have the tendency to bring down the network resources within a span of few minutes. Hence, real-time detection of attacks, especially zero-day exploits, caused by information from untrusted sources is a critical and challenging task.

Several recent works \cite{ptrtaint,lin,song,tainthw,gift,lift,tainttrace,taintbochs,taintsekar,minos} demonstrated that information flow tracking, or \textit{taint analysis} is an effective technique for detecting a wide range of security attacks that corrupt control data such as return addresses or function pointers. This technique works by labeling the input data from untrusted sources as ``unsafe'' or \textit{tainted}. The data derived from such tainted data is marked as tainted as well.  An attack is detected when program control branches to a location specified by the unsafe data e.g., if the tainted data overwrites a function return address. However, due to the fine-grained nature of these techniques i.e., a taint flag per byte of memory, the overhead of taint management and processing is high and slows down the application considerably. Furthermore, the current taint analysis approaches cannot detect \textit{non-control data attacks} \cite{noncontrol}. Consider a case where a buffer is overflowed to overwrite adjacent memory that holds a boolean flag which indicates whether the user has been authenticated or not. By overflowing the buffer and overwriting the boolean flag, an attacker can gain illegitimate access without changing any control data of the application. Since no control data such as return address is overwritten, previous taint analysis based approaches would not be able to detect such an attack. Such non-control data attacks have already been studied in the past \cite{noncontrol}, and past research has shown that the implications of such attacks are same as those of control data attacks, i.e., complete system compromise. Thus, current taint analysis suffer from performance overhead and the inability to detect an important class of attacks aimed at non-control data. 

In this work, we describe coarse-grained object level dynamic taint analysis, a low-overhead information flow tracking technique. Our approach is based on the observation that it is possible to reduce the overhead of taint analysis by  tainting objects i.e., application variables, where each object that has a specific type and size. By doing so, the semantics of the original taint analysis remain unaffected i.e., we will be able track the flow of tainted data in the application. An additional benefit to this coarse-grained approach is that, by knowing the size of the variable we can detect attacks against non-control data by the tainted objects. 

We perform taint analysis on binary executables and hence, do not require the source code. To extract the objects from the executable, we use the symbol table of the application that is generated as part of the debugging information for the executable. Once the variables are extracted we store them along with the executable. Next, we use a user-defined policy to identify untrusted sources of data. Using this policy, when the application is run, we dynamically instrument the application with code that tracks and monitors the data from the untrusted sources. An attack is detected if either the tainted data overwrites control data or if the tainted data tries to overwrite beyond the boundary of an untainted variable. Our contributions are as follows:

\begin{itemize}
\item We present a novel coarse-grained view of taint analysis that preserves the semantics of fine-grained (byte level) taint analysis and reduces the processing overhead. The low computational overhead in our approach makes it feasible for using it in real-time applications. Previously proposed information flow tracking techniques slow down program execution by about 76\% \cite{taintsekar,song,tainttrace,lift}. Our approach reduces this overhead by an order of magnitude, to about 37\%, which indicates that our approach is practical to use in deployed applications. 

\item Our approach is capable of detecting a wide range of attacks including buffer overflows, format string attacks, integer overflows and double-free attacks. Also, some of the previously proposed taint-tracking based techniques do not defend against non-control data attacks \cite{noncontrol}. Our approach can distinguish between different objects, identify their bounds and therefore can defeat non-control data attacks.

\item Unlike some of the previous approaches \cite{taintsekar} that are capable of detecting non-control data attacks, our approach does not require source code of the application. Furthermore, many approaches proposed in the past require special hardware \cite{ptrtaint,tainthw,taintbochs,minos} to propagate taint. Our approach does not require any such non-trivial hardware extensions, and works with existing systems and applications.
\end{itemize}

The rest of the paper is organized as follows. Section \ref{sec:problem} describes some of the most common memory corruption attacks. Section \ref{sec:solution} presents the technical description of our approach. Section \ref{sec:evaluation} presents the experimental results on the effectiveness and performance evaluation. In Section \ref{sec:discussion}, we discuss the ramifications of various security threats and issues in our approach. Finally, in Section \ref{sec:conclusion} we conclude the paper with some future work.

\section{Preliminaries}
\label{sec:problem}
In this section, we introduce the threats being addressed in our work, the related work in this area and our problem statement with assumptions.

\subsection{Security Threats}
Memory corruption vulnerabilities arise when a program uses an unchecked external input. By providing a kind of input that the programmer did not expect, an attacker may cause the program to execute malicious code. The most commonly exploited memory corruption vulnerabilities include buffer overflows and format string bugs. Although, our work addresses integer overflows, heap overflows and a wide range of overflow based attacks, we do not discuss them. They are similar to these popular attacks.

\textbf{Buffer Overflows.}
In a buffer overflow attack, the attacker's aim is to gain access to a system by changing the control flow of a program. Usually, an attacker needs to inject malicious code in the original program execution path. This code can be inserted in the address space of the program using any legitimate form of input. 
To make the program execute this code, the attacker corrupts a code pointer in the program address space by overflowing a buffer and making it point to the injected code.  When the program later dereferences this code pointer, it jumps to the attacker's code. 
Other complicated forms of buffer overflow attacks attempt to change the program control flow by corrupting other code pointers such as, function pointers, global offset table (\texttt{GOT}) entries or longjmp buffers, by overflowing a buffer that may be, locally, globally or dynamically allocated, respectively. Such buffer overflows occur mainly due to the lack of bounds checking in C library functions and laxity of the code writer. For example, the use of \texttt{strcpy()} in a program without ensuring that the destination buffer is at least as large as the source string is a common practice among many $C$ programmers.

\textbf{Format String Attacks.}
Format string vulnerabilities occur when programmers pass user input directly to a format function, such as \texttt{printf()} or \texttt{syslog()}, i.e., using code constructs such as \texttt{printf(str)} instead of \texttt{printf("\%s", str)}. This input is interpreted by the format function as a format string, and is scanned for the presence of format specifiers such as \texttt{\%x}, \texttt{\%s}, \texttt{\%n} etc. For each format specifier, corresponding value or address is picked from the stack and is read or written, depending on the format specifier. To exploit this vulnerability, a common form of the format string attack uses \texttt{\%n} format specifier, which takes an address to an integer as argument, and writes the number of bytes printed so far to the location specified by the address. Using \texttt{\%n} format specifier, an attacker can overwrite the stored return address on the stack with the address of his own code, taking control of the program when the function returns. Other targets include address of destructor functions in \texttt{DTORS} table, address of library functions in the \texttt{GOT}, function pointers and other security critical data.

\textbf{Non-control Data Attacks.}
Classical memory corruption attacks overwrite control data such as stored return addresses, function pointers etc. Such a control-data attack are relatively easily to defend by watching over these control data. However, the same attacks can be made to overwrite program specific non-control data such as a boolean flag that specifies whether the user is authenticated or not. Many real-world software applications are susceptible to non-control data attacks, and the severity of the resulting security compromises is equivalent to that of control-data attacks.

\subsection{Related Work}  
There has been much interest in addressing memory corruption vulnerabilities in the system security community. We classify the related work under three categories.

{\bf Memory Corruption Defenses.} These defenses are mainly aimed to protect the program address space from illegal modifications. Approaches such as StackGuard \cite{stackguard}, StackShield \cite{stackshield}, ProPolice \cite{ssp} and LibSafe \cite{libsafe} and TIED-LibSafePlus \cite{tied} have been proposed as defenses against buffer overflows. Other approaches like FormatGuard \cite{formatguard}, FormatShield \cite{formatshield}, LibFormat \cite{libformat}, Kimchi \cite{kimchi}, Lisbon \cite{lisbon}, White-Listing \cite{whitelist} defend against format string attacks. However, these approaches are limited to certain types of vulnerabilities or exploits, and many of them even require source code of the programs to be protected.

Recently, more general techniques based on randomization of address space have been proposed. The intuition in these approaches is to randomize program address space so that an attacker cannot easily guess the locations where function pointers or return addresses are stored. This has been demonstrated effectively in the Address Space Randomization techniques  described in \cite{aslr,asr2,asr3,aslp} and instruction set randomization in \cite{isr1,isr2}
However, these attacks are susceptible to brute-forcing attacks and can be defeated. Moreover, they are more of a preventive measure in that, the actual buffer overflow vulnerability is only masked by these approaches. 
A determined attacker can still break through such techniques.
Our approach is orthogonal to these approaches as it is  a detection based approach and considers a wider range of attacks.

\textbf{Information Flow Tracking Defenses.} The purpose of information flow tracking is to monitor malicious data as it is processed by the application. At any point if the malicious data violates a user defined policy an attack is said to be found. Information tracking using static taint analysis has been used to find bugs in C programs \cite{shankar,static1,static2} and to find potential sensitive data in Scrash \cite{scrash}. The interpreted language, Perl \cite{perl} does run time taint checking to see whether data from untrusted sources are used in security-sensitive ways such as as argument to a system call. However, the processing overhead is high in Perl and moreover, since it is language based, the applicability is very limited. The approaches described in \cite{taintbochs,tainthw,ptrtaint,minos} perform taint tracing at the hardware level. These techniques are clearly more expensive and difficult to incorporate into existing systems.

Binary instrumentation for taint-tracking was first used in TaintCheck \cite{song}. However, their approach slows down programs significantly (by about $37$x). TaintTrace \cite{tainttrace} achieved significantly faster taint-tracking by using more efficient instrumentation based on DynamoRIO. LIFT \cite{lift} improved the performance benefits further by using better static analysis and faster instrumentation techniques. Although all of these approaches detect a wide range of memory corruption attacks, none of them defends against non-control data attacks, rendering them vulnerable to privilege escalation.

\textbf{Non-control Data Attack Defenses.}
With respect to detecting non-control attacks, the approach in \cite{taintsekar} is the most relevant to our technique compared other taint analysis techniques. However, the approach in \cite{taintsekar} assumes the availability of the application source code, and needs recompilation of the program that is being protected. Therefore, this approach cannot be effective if the source code in unavailable, such as in COTS software or in Microsoft Windows applications. Our approach does not require source code as it uses binary rewriting and dynamic binary instrumentation to monitor information flow.

\subsection{Problem Statement and Assumptions}
We address the problem of detecting memory corruption vulnerabilities that may target control and non-control data. The general problem of thwarting non-control attacks is difficult and requires application context to solve it completely. We look at a subclass of problems that do require minimal application context and can prevent a majority of non-control data attacks. In our approach, we assume that debugging information is present in the executable. This information is used to extract the variables from the symbol table. Note that, this is not a stringent requirement as most software has this information to enable maintenance and repair during its life-cycle. We discuss further about this requirement in section \ref{sec:discussion}. In the next section, we describe our framework in detail.

\section{Object Level Dynamic Taint Analysis}
\label{sec:solution}
This section presents an overview of our approach and the details of the binary instrumentation framework that is required to implement the taint analysis.

\subsection{Approach Overview}
\label{sec:overview}
Our approach is based on the observation that in a program, variables are the entities used by a programmer to accomplish any task, such as receiving user input, reading or writing files, arithmetic, etc. As a program executes, these variables might be copied among themselves (such as \texttt{a=b}), values of some variables might be derived from other variables (such as \texttt{c=a+b}), others might be received from user input (such as \texttt{gets(str)}), etc. An attacker may cause the value of one variable to  illegitimately affect the value of another variable or control data by providing a kind of input that the programmer did not expect, such as an excessively large input, an out of bounds value, an input with format specifiers, etc.

By tracking the source of each variable, we propagate taint on these variables. We refer to these variables in memory as \textit{objects} in our approach. We extract variable addresses and their sizes from the debugging information. By using binary rewriting, this information is inserted in the program binary as a new loadable read-only section, available at run time. At run time, we associate a \textit{tag} with each object, that can have two values: 0 (``safe'')  or 1 (``unsafe'' or ``tainted''). The program is then dynamically instrumented to perform information flow tracking and to detect attacks that alter program control flow to unsafe data. All objects received from untrusted sources, such as network or command line arguments, are marked as tainted. As the program executes, the taint is propagated among objects. Any operation that results in an object being written, causes the tag of the destination object to be affected by that of the source object. Any attempt to branch to an address specified by a tainted object detects an attack. The overview of our approach is shown in Figure 1. Henceforth, we will use the term ``object'' and ``variable'' inter-changeably unless explicitly specified.

\begin{figure}
\centering
\includegraphics[scale=0.5]{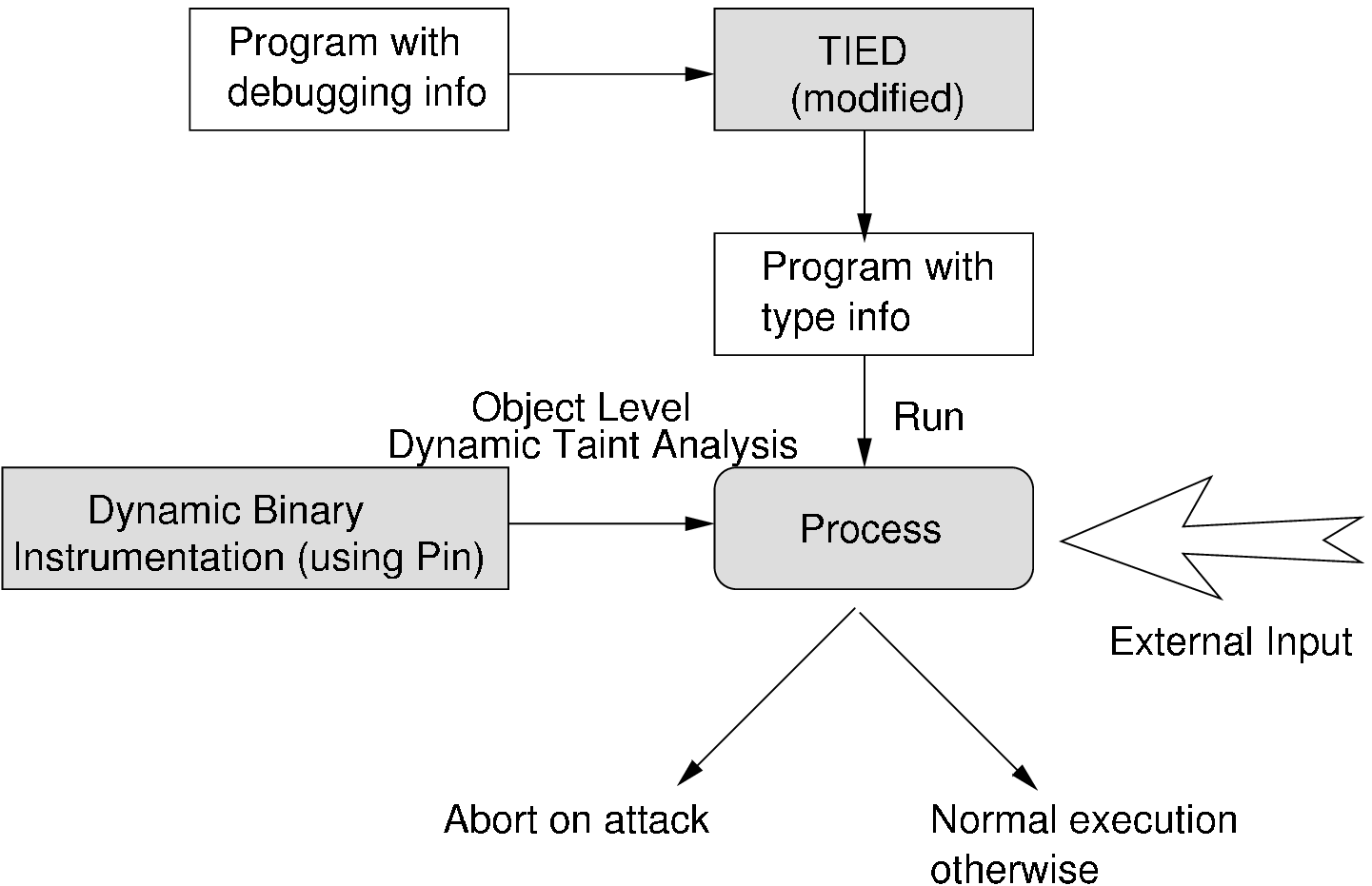}
\caption{Approach Overview}
\end{figure}

{\bf Tainting Policy. } Our tainting policy for objects is simple, if a particular object's state is changed directly by a tainted object then we mark the current object as tainted as well. This can be expressed as follows. If a variable $Z$ is being modified by a function, $F$ that takes one or more tainted objects as parameters, then, $Z$ becomes tainted as well. The function $F$ can be any function that directly affects $Z$. Formally, if $Z$ is affected by $F$ directly i.e., in the same statement then, $Z$ is tainted if one or more parameters of $F$ are also tainted. Note that, this policy excludes implicit flows that often occur due to programmer mistakes, e.g., changing a variable based on logical condition on a tainted value. Moreover, \cite{taintsekar} has noted that tracing implicit flows is a hard problem.

{\bf Justification.	}	\	 Based on our tainting policy, we argue that our approach captures the semantics of the previous taint analysis approaches in \cite{lift,song,tainttrace} and hence, is justified. Earlier approaches propagate taint on every byte of memory. In our approach, note that, the taint tag is propagated on all bytes associated with a single variable. This implies that, even if a single byte of this variable is modified by tainted data, the entire range of bytes associated with the variable become tainted. In fact, our tainting process is a super-set of the byte-level tainting process in \cite{lift,song,tainttrace} as an entire sequence of bytes is tainted even if a single byte is tainted. Furthermore, we propagate taint in all operations where the tainted variables are accessed thereby ensuring that all data affected by tainted data is marked tainted as well. Thus, based on these arguments, our approach appropriately captures the semantics of previous approaches. Later, using experimental evaluation we validate this justification thoroughly.

\subsection{Framework Overview}
Our framework consists of the following steps. First, our approach identifies data objects in the application being checked. We perform static analysis during this phase to determine the instructions that refer to the data objects at run-time. Second, we describe the taint tag management and storage issues in detail. Third, we describe the dynamic instrumentation framework that we use to rewrite the application at run-time. We show the different code instructions that are instrumented at run-time. Finally, we describe the exploit detection policy and the exploits detected using the instrumented code.

\textbf{Identifying Objects.}
The first step in our approach is to extract the objects from the executable and make it available to our taint tracking framework at run-time. Note that, for each executable, this is a one-time process. The extracted information is added to the executable and stored along with it. 
Our object extraction uses an enhanced version of the TIED\footnote{Type Information Extractor and Depositor} \cite{tied} framework which is suitable for our purpose. Since the program is compiled with debugging information, it contains variables location and size information that can be used at run time. TIED extracts location and size information of all the buffers in the program and dumps it in the executable as a separate loadable read-only section. 
For global variables, the addresses are known at compile time itself, while for local variables only offset from the frame pointer is known at compile time. We have modified TIED so that it dumps the location (virtual address for the global variables and offset from frame pointer for local variables) and size information of all the variables in the executable. The members of arrays, structures and unions are also explored to detect the variables that lie in them. Figure 2 demonstrates a typical case of variables within structures. TIED detects all 40 variables in this case. This location and size information is loaded at run time to identify objects in the virtual address space of the program. 
Figure 3 shows a code fragment and the corresponding objects that are created at run time. For the example code, the previous approaches would read the input up to 1024 bytes, mark the tag for all the bytes read as tainted and then propagate the taint to the destination. However, in our approach we mark the object \texttt{src} as tainted and taint is propagated to \texttt{dest} object. Therefore, for the same code, our approach marks an entity as tainted only once and taint propagation is required just once, thus saving taint marking as well as propagation effort to a large extent.\\
\begin{figure}
\centering
\mbox{
\begin{lstlisting}
struct s {
	char a;
	int b;
};
struct s foo[20];
\end{lstlisting}}
\caption{Variables within a structure}
\end{figure}
\begin{figure}
\centering
\begin{minipage}{3in}
\mbox{
\begin{lstlisting}
void readdata(int fd) {
	char dest[1024];
	char src[1024];
	
	read(fd, src, sizeof(src));
	strcpy(dest, src);
}	
\end{lstlisting}}
\end{minipage}
\qquad
\begin{minipage}{3in}
\includegraphics[scale=0.5]{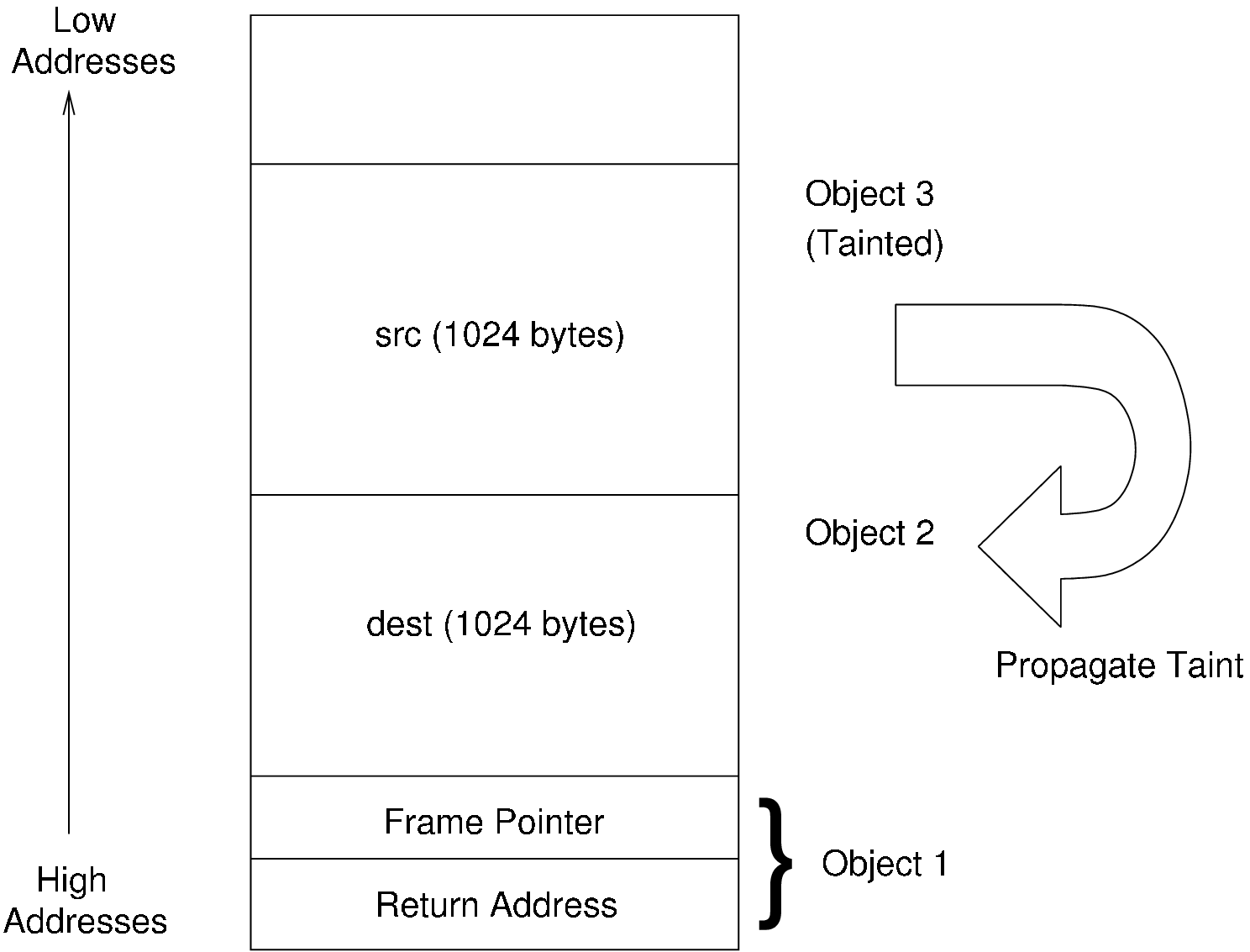}
\end{minipage}
\caption{Objects in a stack frame}
\end{figure}
{\bf Static Analysis.	} \	We statically analyze the program to identify instructions referring to these variables in memory at program run-time. Global variables are identified by looking at the address being referred by the instruction. Local variables and function arguments are identified by looking at the offset from frame pointer. 
A negative offset identifies a local variable and a positive offset identifies a function argument.

\textbf{Tag Management.}
We associate a one bit tag with each object and register, which can have only two values (0 for ``safe'' and 1 for ``unsafe''). 
As a single bit tag is maintained for an object that may span several bytes, the space overhead is considerably lower than approaches that use a single-bit tag per byte of memory. Instead of maintaining tags for the entire program's address space, we maintain tags only for writable parts of the program. This is done since the non-writable parts of the address space cannot be affected by the user input and are therefore assumed to be ``safe''. Hence, the tag for any object belonging to a read-only part of the address space is assumed to be 0 or ``untainted''.

We store tags for objects in a separate memory region, called \textit{tag space} (or \textit{shadow memory}), which is addressable via a one-to-one direct mapping between an object and the tag-bit in the program's virtual address space. Such a direct mapping makes it straightforward and fast (with only one memory access and a few arithmetic instructions) to get the tag value for a given object. At program load time, our prototype uses the information dumped by TIED to identify objects in the process address space. For every object, our approach maintains a (\texttt{offset}, \texttt{bit}) tuple, which is used to locate its tag bit. The \texttt{offset} refers to the difference between the tag byte that holds the tag bit for an object and the beginning of the tag space, and the \texttt{bit} holds the bit position in that byte. This tuple is determined using static analysis of the program during load time. At program load time, instructions that refer to an object are located, and are instrumented so as to propagate taint at runtime. For local variables, for which memory is reserved on the stack only when the corresponding function is called, this step is deferred till the corresponding function is called.

\textbf{Dynamic Binary Instrumentation Framework and Tracking.}
To enable taint tracking, when the program is executed, we dynamically instrument the code. The instrumentation process supplements the code with additional instructions to enable object tracking and taint propagation. The instrumentation needs to be done every time the program is restarted. Our dynamic binary instrumentation approach is built on top an existing dynamic binary instrumentation framework called Pin \cite{pin}. Pin provides efficient instrumentation by using a just-in-time (JIT) compiler to insert and optimize code. In addition to some standard techniques for dynamic instrumentation systems including code caching and trace linking. Pin implements register reallocation, liveliness analysis, code inlining and instruction scheduling to optimize jitted code. 

A tag space of one bit is maintained for each object that was identified during static analysis.
At program initialization, all tags are cleared to zero. Data (objects) received from untrusted sources such as network or standard input are marked as tainted. As the program executes, other objects may be tagged as tainted via information flow. A tainted object may become untainted if its value is reassigned from some untainted object. In accordance to our formal definition of {\em taint policy} (cf. Section \ref{sec:overview}), we instrument all data movement and arithmetic instructions.

\begin{itemize}
\item For data movement instructions such as \texttt{mov}, \texttt{lods}, \texttt{stos}, \texttt{push}, \texttt{pop}, etc, the tag value of the source operand is propagated to that of the destination. For data movement instructions which move a constant value in an object, we instrument the instruction to mark the object as untainted at run time.
\item For arithmetic instructions, such as \texttt{add}, \texttt{sub}, \texttt{xor}, etc, the tag value of the destination is the OR of the tag values of the source operands, since the value of the destination is affected by both the source operands.
\item For instructions which do not involve any second operand (explicit or implicit), such as \texttt{inc}, \texttt{dec}, etc, the tag value does not change.
\end{itemize}
\begin{figure}
\centering
\mbox{
\begin{lstlisting}
movb $0x10,%eax
mov  $0xfffffffe,%ecx
rol  $0x03,%ecx
and  %ecx,$0xa8000000(%eax)
mov  $0x0400,%ebx
\end{lstlisting}}
\caption{Instrumentation for the instruction \texttt{mov \$0x0400,\%ebx}}
\end{figure}
In Figure 4, we show the instrumentation code generated for the instruction \texttt{mov \$0x0400,\%ebx}. As this instruction is associated with moving data it needs to be instrumented. Since a constant value \texttt{0x0400} is being copied to \texttt{ebx} register, the register needs to be marked as untainted. To do that, a value (\texttt{0x10} in this example) is copied to register \texttt{eax}. This value is an offset from the beginning of the tag space (\texttt{0xA8000000} in this example). Another value \texttt{0xfffffffe} is copied to \texttt{ecx} register. This value is then rotated left by 3 bits to create a mask, which is then ANDed with the byte at address \texttt{A800000000 + 0x10 = 0xA800000010}, to mark the corresponding taint bit for the \texttt{ebx} register as untainted. The original instruction is then executed. 

Other instructions are identified using the rules described above. Note that, these rules are not applicable for a few special instructions for which the result is untainted irrespective of whether the operand is tainted or not. For example, in x86 architecture, the instructions such as ``\texttt{xor \%eax, \%eax}'' or ``\texttt{sub \%eax, \%eax}'' clear the \texttt{eax} register irrespective of its previous value. In our approach, we identify such instructions and mark the corresponding register as untainted. We instrument calls to functions that allocate memory such as \texttt{malloc()} to identify corresponding object in global shadow memory. Calls to functions that copy memory such as \texttt{memcpy()} or \texttt{strcpy()} are instrumented to propagate taint from the source object to the destination object. Calls to functions such as \texttt{memset()} are instrumented to mark the corresponding argument as untainted if the source byte lies in an untainted object. For \texttt{mmap} or \texttt{mmap2} system calls, we check the return value to see if the allocation was successful. For each successful allocation, we identify objects in the global shadow memory. In Table 1, we show a categorized list (partial) of instructions that are identified for instrumentation.
\begin{table}
\caption{Instructions instrumented for Taint Tracking}
\centering
\begin{tabular}{|p{60mm}|p{60mm}|} \hline \hline
\textbf{Instruction type / System or library call} & \textbf{Instrumentation} \\ \hline
Data movement instructions, \textit{e.g.} \texttt{mov}, \texttt{lods}, \texttt{stos}, etc. & tag[dest] = tag[source] \\ \hline
Arithmetic instructions, \textit{e.g.} \texttt{add}, \texttt{sub}, \texttt{xor}, etc. & tag[dest] = tag[a] OR tag[b] \\ \hline
Instructions without any second operand, \textit{e.g.} \texttt{inc}, \texttt{dec}, etc. & No change in tag value \\ \hline
Calls that copy memory, \textit{e.g.} \texttt{memcpy}, \texttt{strcpy}, etc. & tag[dest] = tag[source] \\ \hline
Calls that set memory, \textit{e.g} \texttt{memset} & tag[dest] = tag[source] if source is not constant, else tag[dest] = 0 \\ \hline \hline
\end{tabular}
\end{table}

\subsection{Exploit Detection Policy}
We use a configurable exploit detection policy to detect any exploitation attempts. The exploit detection policy specifies the sources that are untrusted, for example, network, files, standard input, command line arguments, environment variables, etc. Any data object received from an untrusted source is marked as tainted. The policy also defines the checks that are applied when the program is run. Based on our exploit detection policy, we instrument all control altering instructions such as \texttt{call}, \texttt{jmp}, \texttt{ret}, etc. and check the branch address. A branch address belonging to a tainted object detects a control hijacking attack. To detect format string attacks, we instrument calls to functions in the \texttt{printf}, \texttt{syslog}, and \texttt{warn/err} family. A format string belonging to a tainted object detects a format string attack. The policy also checks whether any non-control objects are marked as tainted in an illegitimate manner. For example, a stack overflow attack that does not overwrite the stored return address, but only overwrites the next object on the stack is noted in the exploit detection policy. Table 3 lists the exploit detection policies that are required for detecting different types of attacks.
\begin{table}
\caption{Attacks detected by different policies}
\centering
\begin{tabular}{|p{60mm}|p{60mm}|} \hline \hline
\textbf{Policy} & \textbf{Attacks Detected}\\ \hline
Tainted branch addresses & Classical stack smashing attacks, common buffer overflows and frame faking attacks\\ \hline
Tainted format string arguments & Format string attacks\\ \hline
Tainted system call arguments & Any kind of code injection attacks\\ \hline
Tainted control data, \textit{e.g.} tainted boundary tags in heap or \texttt{longjmp} buffers etc. & Heap overflows and stack overflows\\ \hline
Tainted Non-control data, \textit{e.g.} variable containing user privileges etc. & All kinds of non-control data attacks\\ \hline \hline
\end{tabular}
\end{table}

The following discussion illustrates the attacks and the checks that are enforced by our exploit detection policy for detecting these attacks.
\begin{itemize}
\item \textit{Branch Addresses.} The policy checks if any address is used as a target of any branch instruction such as \texttt{jmp} or \texttt{call}. This defeats all those attacks which overwrite return addresses, function pointers, global offset table (\texttt{GOT}) entries, etc. in order to redirect the program control flow to the injected code or to some library function such as \texttt{execve()}. It also defeats \textit{frame faking attacks} that overwrite the stored frame pointer in order to create a fake stack frame when the function returns. 
\item \textit{Format string arguments.} The policy checks if any format function such \texttt{fprintf()} or \texttt{syslog()} is called with a tainted format string argument. Any such case is identified as an exploitation attempt irrespective of the format specifiers used. Therefore, it can not only defeat attempts to overwrite arbitrary memory locations, but even arbitrary memory read attempts. Note that such a check can not only detect an exploitation attempt, but can also pinpoint the vulnerability even before it is exploited. This is because the format string argument will still be tainted even if the input is legitimate. 
\item \textit{System call arguments.} The policy checks if any arguments used by a system call are overwritten by tainted data. In Linux x86 architecture, a system call is implemented using software interrupt 0x80. The system call number is usually passed in the \texttt{eax} register. We instrument all \texttt{int 0x80} instructions to check the value of \texttt{eax} register at run time. For \texttt{execve} system call, we check at run time if its arguments are tainted. Such checks will not only defeat code injection attacks, but even those attacks in which an attacker overwrites data that is later used as an argument to a system call. 
\item \textit{Other control data.} The policy also checks whether certain other control data such as boundary tags in heap or \texttt{longjmp} buffer are tainted. Such tags are inserted at points where these control data are used. For example, for boundary tags in heap, we insert checks in calls to \texttt{free()}, and for \texttt{longjmp} buffers, we insert checks in calls to \texttt{longjmp()} and \texttt{siglongjmp()}.
\item \textit{Non-control data.} The policy checks if any non-control data is marked as tainted in an illegitimate manner, such as by an overflow of an object. Such checks are implemented by checking bounds of the object being written, as done by previous approaches such as LibSafe \cite{libsafe} or LibSafePlus \cite{tied}.
\end{itemize}

In the next section, we evaluate the effectiveness of our taint analysis approach against these attacks. We validate our technique through extensive testing over synthetic and real-time exploits.

\section{Evaluation}
\label{sec:evaluation}
We have implemented a prototype of our approach on Linux x86 architecture. We use TIED to dump variable size and location information into the program executable. Our prototype then uses Pin to instrument the code by adding instructions to perform information flow tracking on these objects and to detect attacks. The instrumentation is performed only once, at the program load time. However, the instrumented code may run many times. Pin caches the generated instrumented code, so that it need not be instrumented again when required.

We conducted a series of experiments to evaluate the effectiveness and performance of our approach. All tests were run in single user mode on a Pentium-4 3.2 GHz machine with 512 MB RAM running Linux kernel 2.6.18. All programs were compiled with gcc 4.1.2 and linked with glibc 2.3.6.

\subsection{Effectiveness Evaluation}
We tested the effectiveness of our approach using several synthetic and actual exploits. The exploits were selected from a wide range of attacks including stack smashing attack, heap overflow, format string attack, double free and non-control data attacks. Our approach successfully detected all the exploitation attempts and terminated the victim program to prevent execution of malicious code. The results are presented in Table 3.
\begin{table}[!h]
\caption{Results of effectiveness evaluation}
\centering
\begin{tabular}{|c|c|c|c|p{30mm}|c|} \hline \hline
\textbf{CVE\#} & \textbf{Program} & \textbf{Attack Type} & \textbf{Overwrite Target} & \textbf{Description} & \textbf{Detected}\\ \hline
- & Synthetic & Stack Smashing & Stored return address & Stack Smashing Attack & \checkmark \\ \hline
- & Synthetic & Heap Overflow & Boundary tags & Heap Overflow Attack & \checkmark \\ \hline
- & Synthetic & Format String Attack & GOT entry & Format String Attack & \checkmark \\ \hline
- & Synthetic & Buffer Overflow & Array in a structure & Non-control data attack & \checkmark \\ \hline
CVE-2003-0201 & samba & Buffer Overflow & Stored return address & Buffer overflow in \texttt{call\_trans2open()} function & \checkmark \\ \hline
CVE-2001-0010 & bind & Buffer Overflow & Stored return address & Buffer overflow in \texttt{nslookupComplain()} routine & \checkmark \\ \hline
CVE-2000-0763 & xlock & Format String Attack & Stored return address & Format string vulnerability when using \texttt{-d} command line switch & \checkmark \\ \hline
CVE-2007-1711 & PHP & Double free attack & Pointer to destructor & Double free vulnerability using \texttt{session\_decode()} function & \checkmark \\ \hline
CVE-2007-4566 & SIDVault & Buffer Overflow & Stored return address & Buffer Overflow in \texttt{Simple\_Bind()} function & \checkmark \\ \hline
CVE-2001-0820 & ghttpd & Buffer Overflow & Stored \texttt{ESI} register & Non-control data attack & \checkmark \\ \hline
CVE-2000-0573 & wuftpd & Format String Attack & Cached copy of user ID & Non-control data attack & \checkmark \\ \hline \hline
\end{tabular}
\end{table}
\subsubsection{Synthetic Exploits.}
In this section, we evaluate our approach using synthetic exploits on stack overflows, heap overflows, and format string vulnerabilities. Our approach successfully detected all the attacks without any false negatives.
\begin{itemize}
\item \textit{Stack Smashing Attack.}
To test the effectiveness of our approach against the classical stack smashing attack, we wrote a program with a buffer overflow vulnerability. It uses \texttt{strcpy()} to copy a command line argument to a local buffer. We tried inserting a long command line argument in order to overflow the buffer and to overwrite the stored return address. Our approach successfully detected the tainted return address (which is treated like any other control data in our approach) when the function returned.
\item \textit{Heap Overflow Attack.}
In a similar test, we verified our approach by detecting a heap overflow. We wrote a program with a heap overflow vulnerability. It allocates a buffer in heap, uses \texttt{strcpy()} to copy the first command line input into the buffer, and then frees the buffer. We injected a huge input in the first command line argument in order to overflow the buffer and overwrite the boundary tags. Our approach successfully detected the tainted boundary tags when the buffer was freed.
\item \textit{Format String Attack.}
We used our approach on a custom written program with a format string vulnerability. The program uses \texttt{printf()} to display the first command line argument. The exploit for the program uses the format string attack to overwrite the global offset table (\texttt{GOT}) entry for \texttt{exit()} function. The exploitation attempt was successfully detected as the format string argument was found to be tainted. Note that for a format string attack, our approach can defeat all exploitation attempts of writing to and reading from arbitrary memory locations irrespective of the format specifier used.
\item \textit{Buffer Overflow Attack.}
To test our approach against non-control data attacks, we tried overflowing a buffer in a structure in order to overwrite a character array stored next to it. The character array is used to hold the filename of the temporary file being written. The attack was detected by the fine grained policy as the character array was found to be illegitimately written by a tainted buffer.
\end{itemize}

\subsubsection{Actual Exploits.}
We tested our approach on five exploits on publicly known vulnerabilities. All the attacks were successfully detected by our approach.
\begin{itemize}
\item \textit{Samba \texttt{call\_trans2open()} Buffer Overflow.}
Samba version 2.2.8 and earlier suffer from a stack smashing vulnerability \cite{smbbuf} in \texttt{call\_trans2open()} function. Successful exploitation of the vulnerability allows a remote attacker to gain a root shell on the machine running vulnerable version of samba. Our approach detected the tainted stored return address and defeated the exploitation attempt.
\item \textit{BIND 8 Buffer Overflow.}
BIND version 8.2 and earlier suffer from a buffer overflow \cite{bindvul} in the \\ \texttt{nslookupComplain()} routine, which allows a remote attacker to gain root access on the affected machine. Our approach correctly detected that the return address is tainted and defeated the attack.
\item \textit{xlockmore Format String Vulnerability.}
xlock version 4.16 suffers from a format string vulnerability \cite{xlock} when using the command line argument \texttt{-d}, that can be used by a local user to gain root privileges. The exploit overwrites the stored return address with the address of the injected shellcode. Our approach successfully identified that the stored return address was tainted and defeated the exploitation attempt.
\item \textit{PHP \texttt{Session\_Decode()} Double Free Memory Corruption Vulnerability.}
PHP version 4.4.5 and 4.5.6 suffer from a double free vulnerability \cite{phpdoublefree}, that can be used by a local user to execute arbitrary code in the context of the webserver or to cause denial of service conditions. The exploit overwrites the pointer to a destructor with a junk value to cause denial of service. Our approach successfully identified that the pointer to destructor was tainted and defeated the exploitation attempt.
\item \textit{ghttpd \texttt{Log()} Buffer Overflow Vulnerability.}
A stack overflow vulnerability \cite{ghttpd} exists in ghttpd version 1.4.3 and lower which allows a remote user to execute arbitrary code with the privileges of the web server. The overflow occurs in \texttt{Log()} when the argument to a GET request overruns a 200-byte stack buffer. In order to test our approach against non-control data attacks, we modified the publicly available exploit to overwrite the stored \texttt{ESI} register on the stack, which is a later copied to a pointer to the URL requested by the client. By overwriting the stored \texttt{ESI} register, it is made to point to the URL \texttt{/cgi-bin/../../../../bin/sh} in order to execute a shell. Our approach correctly detected the argument to the \texttt{execlp()} was tainted and defeated the attack.
\item \textit{wuftpd \texttt{SITE EXEC} Format String Vulnerability.}
wuftpd version 2.6.0 and earlier suffer from a format string vulnerability \cite{wuftpd} in \texttt{SITE EXEC} implementation that allows arbitrary code execution. We modified the publicly available exploit for this vulnerability to overwrite the cached copy of user ID \texttt{pw->pw\_uid} with 0 so as to disable the server's ability to drop privileges. The attack was detected as the format string argument was found to be tainted.
\item \textit{SIDVault \texttt{Simple\_Bind()} Function Remote Buffer Overflow Vulnerability.}
SIDVault LDAP server version 2.0e and earlier suffer from a buffer overflow \cite{sidvault}, that can be used by a remote user to gain root privileges. The exploit overwrites the stored return address with the address of the injected shellcode. Our approach successfully identified that the stored return address was tainted and defeated the exploitation attempt.
\end{itemize}

\noindent{\bf Note on SIDVault.} We would like to mention here that our approach was tested on SIDVault LDAP server buffer overflow attack \cite{sidvault}, which is a proprietary software available without source code. Whereas our approach is applicable to such proprietary software available without source code, other approaches that rely on source code such as Xu et al's approach \cite{taintsekar} will not be able to defend against attacks in such cases.

\subsection{Performance Evaluation}
To test the performance overhead of our approach, we used our approach on several CPU intensive programs:
\begin{itemize}
\item \textbf{bc} - bc is an interactive algebraic language with arbitrary precision, with several extensions including multi-character variable names, and full Boolean expressions. The test was to calculate the factorial of 600.
\item \textbf{Enscript} - Enscript converts ASCII files to PostScript, HTML, RTF or Pretty-Print and stores generated output to a file or sends it directly to the printer. The test was to convert a 5.5 MB text file to postscript.
\item \textbf{Bison} - Bison is a general-purpose parser generator that converts a grammar description for an LALR(1) context-free grammar into a program to parse that grammar. The test was to parse a bison file for C++ grammer.
\item \textbf{gzip} - Gzip is the standard file compression utility. The test was to compress a 12 MB file.
\end{itemize}
We performed all the tests 10000 times and took the averages of the results. We compare our results to that of Xu et al's approach, which has the lowest overhead among all the previous approaches. Our approach incurred an average overhead of 37.2\% while that of Xu et al's approach \cite{taintsekar} was 76\%, which clearly shows an improvement by an order of magnitude. We present the test results in Table 4 and in Table 5 we present a detailed feature comparison of our approach with previous approaches.
\begin{table}
\caption{Comparison of overhead incurred by our approach with that of Xu et al's approach}
\centering
\begin{tabular}{|l|l|c|c|} \hline \hline
~ \textbf{Program} & ~~~~~~~~~~~~ \textbf{Test} & \multicolumn{2}{ c |}{\textbf{Overhead}} \\ \cline{3-4}
 & & \textbf{Xu et al's} & \textbf{Our} \\
 & & \textbf{approach \cite{taintsekar}} & \textbf{approach} \\ \hline
bc-1.06 & Find factorial of 600 & 61\% & 42.84\% \\ \hline
enscript-1.6.4 & Convert a 5.5 MB text file into a PS file & 58\% & 28.63\% \\ \hline
bison-1.35 & Parse a Bison file for C++ grammer & 78\% & 32.02\% \\ \hline
gzip-1.3.3 & Compress a 12 MB file & 106\% & 45.38\% \\ \hline \hline
\end{tabular}
\end{table}
\begin{table}
\caption{Feature Comparison with previous approaches}
\centering
\begin{tabular}{ | l | c | c | c | p{14mm} | p{17mm} | p{14mm} | }
\hline \hline
\textbf{Feature} & \textbf{TaintCheck \cite{song}} & \textbf{TaintTrace \cite{tainttrace}} & \textbf{LIFT \cite{lift}} & \textbf{Chen et al's approach \cite{ptrtaint}} & \textbf{Xu et al's approach \cite{taintsekar}} & \textbf{Our approach} \\ \hline
Works without additional hardware & \checkmark & \checkmark & \checkmark & $\quad$ \ding{53} & $\quad$ \checkmark & $\quad$ \checkmark \\ \hline
Works without source code & \checkmark & \checkmark & \checkmark & $\quad$ \checkmark & $\quad$ \ding{53} & $\quad$ \checkmark \footnotemark[2] \\ \hline
Detects non control data attacks & \ding{53} & \ding{53} & \ding{53} & $\quad$ \checkmark \footnotemark[3] & $\quad$ \checkmark & $\quad$ \checkmark \\ \hline
Performance Overhead & 3720\% & 553\% & 360\% & $\quad$ - \footnotemark[4] & $\quad$ 76\% & ~~37.22\% \\ \hline \hline
\end{tabular}
\end{table}
\footnotetext[2]{Requires debugging information}
\footnotetext[3]{Can only detect non-control data attacks where a pointer is tainted}
\footnotetext[4]{Performance results were not available}

\section{Discussion}
\label{sec:discussion}
In this section, in the context of our approach, we discuss ramifications of different security threats and address concerns regarding protection of the shadow memory.

\textbf{Ramifications of Varied Security Threats.} In the present day scenario, attacks on web applications, such as XSS and SQL injection, do contribute to the majority of the attacks today, however, the impact of such attacks is limited than that of memory corruption attacks. For e.g., for a server having both an XSS vulnerability and a buffer overflow vulnerability, exploitation of XSS flaw may lead to credit card details of a user being stolen, while exploitation of buffer overflow may lead to credit card details of all the users being stolen in a single attempt. Furthermore, exploitation of XSS would give an attacker only a limited (user-level) access on the server while exploitation of buffer overflow would give him superuser access (typically) on the server and it would even give him access to the corporate intranet. It is important to note that, memory corruption attacks have been used by worms such as CodeRed and Slammer, which plague the Internet in a matter of few minutes. Defending from web application attacks is easier due to the availability of the source code. However defending from memory corruption attacks is not easy due to source code unavailability, performance considerations and other factors. Construction of exploits for memory corruption attacks requires high expertise and experience. However, construction of exploits for XSS and SQL injection is fairly easy. This is easily justified by the fact that a single zero-day exploit for a Microsoft platform costs between 10000 - 100000 USD, in the underground market. However, several freeware tools are available for exploiting XSS and SQL injection vulnerabilities.

\textbf{Protection of Shadow Memory.} For protecting the shadow memory we used an approach similar to that used in \cite{lift}. Our shadow memory mapping strategy can prevent attackers from directly modifying shadow memory as long as the attackers' code is under the control of Pin. The reason is that when attackers issue an instruction to access an address \texttt{addr} in the shadow memory, the instrumented code will access memory at \texttt{addr+shadow\_base}, which is beyond the boundary of the shadow memory area. This will cause an invalid memory access exception. Note that an attacker might try to access the shadow memory by using a large value of \texttt{addr} so that \texttt{addr+shadow\_base} overflows the 32-bits of the integer used to hold the sum. To counter such attacks, we check if the sum \texttt{addr+shadow\_base} leads to an overflow condition.


\textbf{Implicit Information Flows.}
Implicit information flow occurs when the values of certain variables are related by virtue of program logic, even though there are no assignments between them. For example, the following code snippet illustrates such a case.
\begin{center}
\texttt{if(x==0) y=0; if(x==1) y=1;}
\end{center}
The above code fragment is equivalent to the assignment \texttt{y=x}. Therefore, if \texttt{x} is tainted (untainted), \texttt{y} is tainted (untainted) as well, although there is no direct assignment from \texttt{x} to \texttt{y}. But since the value of \texttt{y} is assigned from a constant, it would be marked as untainted, irrespective of the tag of \texttt{x}. Our approach supports some of such basic implicit information flows, such as the use of translation tables, which are sometimes used for decoding using table lookups, such as \texttt{y = table[x]}. In such a case, the array index \texttt{x} determines the value of \texttt{y}. To handle such cases, our approach marks the result of an array access as tainted (untainted) whenever the index is tainted (untainted). Much more complex implicit flows can be supported using a notion of noninterference, which however is too conservative and leads to high false positives.

\textbf{Limitations.}
Our approach is built on top of TIED. Although TIED does not require source code of the programs, it requires them to be compiled with debugging information. A limitation of our approach is that it does not work for stripped binaries, i.e. programs without debugging and other symbol table information. However, we believe it is not such a major limitation as many operating systems distributions, for example Microsoft Windows and Fedora family of operating systems make debugging information available as separate packages \cite{msdebug,fedoradebug}. Requirement for debugging information may also ease reverse engineering in some cases. However, we argue that our approach requires only information related to the variables. The debugging information related to code can safely be removed from the program binaries. Even in the presence of debugging information, obfuscations can still be applied to the code present in the binary to make it harder to reverse engineer. Moreover, some approaches \cite{memaccess,variables} have been proposed in the past that are able to discover variables in stripped binaries with a high accuracy. We believe using a similar technique would make our approach independent of the debugging information present in the executables, making it work even for stripped binaries.

\section{Conclusion and Future Work}
\label{sec:conclusion}
In this paper, we presented a novel coarse-grained taint analysis approach to detect memory corruption attacks. Our approach propagates taint on data objects that are extracted from the debugging information of the application. Attacks are detected when a tainted object is used in a security-sensitive operation or when a tainted control-data is used. Our evaluation showed a substantial improvement in performance over that reported by previous works on taint tracking. Also, our approach detected a wide range of memory corruption attacks, including non-control data attacks. 

One of the future works in our approach involves extracting variable information effectively without the need for debugging information.  Also, we are exploring ways of incorporating more application context information so as to detect other attacks on critical variables that are identified by the application developer. 
\bibliographystyle{unsrt}
\bibliography{main}

\end{document}